\documentclass[10pt,conference]{IEEEtran}

\usepackage{cite}

\usepackage{rotating}
\usepackage{multirow}
\usepackage{graphicx}
\usepackage{epsfig}
\usepackage{latexsym}
\usepackage{url}
\usepackage{amssymb,amsmath}
\begin{document}

\title{Stuck in Traffic (SiT) Attacks:\\{\Large A Framework for Identifying Stealthy
Attacks that Cause Traffic Congestion}}

\author{\authorblockN{\sc{Mina Guirguis}}
\authorblockA{Dept. of Computer Science\\
                  Texas State University\\
                  \tt{msg@txstate.edu}}
\and
\authorblockN{\sc{George Atia}}
\authorblockA{Dept. of Electrical Engineering and Computer Science\\
                  University of Central Florida\\
                  \tt{george.atia@ucf.edu}} }

\maketitle

\begin{abstract}
Recent advances in wireless technologies have enabled many new
applications in Intelligent Transportation Systems (ITS) such as
collision avoidance, cooperative driving, congestion avoidance,
and traffic optimization. Due to the vulnerable nature of wireless
communication against interference and intentional jamming, ITS
face new challenges to ensure the reliability and the safety of
the overall system. In this paper, we expose a class of stealthy
attacks -- Stuck in Traffic (SiT) attacks -- that aim to cause
congestion by exploiting how drivers make decisions based on smart
traffic signs. An attacker mounting a SiT attack solves a Markov
Decision Process problem to find optimal/suboptimal attack
policies in which he/she interferes with a well-chosen subset of
signals that are based on the state of the system. We apply
Approximate Policy Iteration (API) algorithms to derive potent
attack policies. We evaluate their performance on a number of
systems and compare them to other attack policies including
random, myopic and DoS attack policies. The generated policies,
albeit suboptimal, are shown to significantly outperform other
attack policies as they maximize the expected cumulative reward
from the standpoint of the attacker.

\end{abstract}

\section{Introduction}\label{sec:intro}
In the area of Intelligent
Transportation Systems (ITS), vehicle-to-vehicle and
vehicle-to-infrastructure communications enable many potential
applications such as collision avoidance, cooperative driving,
congestion avoidance, and traffic optimization
\cite{Elbatt06cooperativecollision,misener2005cooperative,yan2010cooperative,tatchikou2005cooperative,biswas2006vehicle}.
With recent advances in wireless technologies, the FCC has
allocated a frequency band of about 75MHz in the 5.9 GHz band for
Dedicated Short Range Communications (DSRC) for public safety
services \cite{FCC}. The associated MAC layer can be based on WLAN
(IEEE 802.11p) or 3G cellular extended with TDMA and CDMA for
decentralized access when no infrastructure is present
\cite{luosurvey,sichitiu2008inter}.\vspace{0.03in}

Due to the shared nature of the wireless channels used, the
overall safety of the ITS is affected by interference and
intentional jamming by adversaries. Jamming has been shown to
cause severe effects that may cripple the whole system
\cite{jamming,thuente2006intelligent,sang2009capabilities,reactivejamming2011,pelechrinis2011denial,law2009energy}.
Previous incidents indicate the possibility of interfering with
these communication mediums \cite{diymedia}. By placing jamming
devices in vehicles and at critical transportation points
(bridges, tunnels, cellular towers, etc...), an adversary can
impact the overall traffic flow, exploit the adaptation of the
drivers to make abrupt decisions causing accidents, or attempt to
maximize their gain by preventing critical information from
reaching a neighboring subset of vehicles \cite{raya,aijaz}. A
much worse scenario may occur if a terrorist can create severe
congestion in an area before detonating a bomb.\vspace{0.03in}

As drivers increase their reliance on wireless signals in making
decisions, the absence or even the delay of these signals may have
catastrophic effects due to the real-time constraints present in
the system. Although it may be still years for autonomous vehicles
to become the main stream (with all the challenges currently
present), nowadays, we rely on real-time traffic information to
make driving decisions. Moreover, many vehicles are already
equipped with wireless connections to invoke traffic
services.\vspace{0.03in}

\noindent {\bf Paper scope:} Drivers are typically faced with a
decision making process whenever they encounter alternatives in
choosing their routes. For example, should a driver use the upper
or lower level when driving across George Washington Bridge?
Should a driver use a highway or a local access road for a given
short trip? The decisions made are not random, but are typically
aided by traffic signs (e.g., reflecting the delay or the expected
time to reach a particular point) and/or online map services
(e.g., Google maps with traffic information). The goal is to
reduce congestion as much as possible. It is known that traffic
congestion is a significant problem that costs the US billions of
dollars. In 2010, and across 439 urban areas, traffic congestion
came at the price of 4.8 billion hours of extra driving time and
1.9 billion gallons of fuel. The cost to the average commuter was
\$713 in 2010 \cite{mobilityreport}.\vspace{0.03in}

When wireless signals are used to communicate important
information to drivers -- perhaps through smart traffic signs and
wireless transceivers in vehicles -- jamming a subset of the
signals may impact the overall traffic flow leading to unchecked
safety conditions. In this paper, we expose a class of stealthy
attacks -- that we term Stuck in Traffic (SiT) attacks -- that aim
to cause congestion. Through solving a Markov Decision Process
(MDP) problem, an attacker mounting a SiT attack selects a subset
of signals to interfere with. The choice of signals is based on
the current state conditions of the system, taking into account
the exposure risk the attacker is willing to take. Due to the
exponential nature of the state space that describes the system,
solving the MDP exactly is computationally prohibitive. Thus, we
apply approximate policy iteration methods to solve the MDP
problem to identify suboptimal, yet efficient, attack
policies.\vspace{0.03in}

\noindent {\bf Contributions:} The use of wireless technologies in
various traffic safety applications is becoming the norm. Thus, it
is important to expose potential security issues before
deployment. In particular, we make the following contributions:

\begin{itemize}
    \item We provide a general framework for identifying stealthy attacks
    that reflect the best interest for an attacker: minimizing the
    cost while maximizing the damage.
    \item We expose SiT attacks that aim to cause traffic congestion
    through a proper choice of {\it which} signals to interfere with and
    {\it when}.
    \item  In almost all the cases studied, we were able to identify attack
    policies that are more potent than traditional DoS attacks and random
    attacks, among other policies. Furthermore, the generated policies are shown to significantly outperform myopic and random attack policies.
\end{itemize}

\vspace{0.03in} \noindent {\bf Paper organization:} In Section
\ref{sec:relwork}, we survey related work. In Section
\ref{sec:model}, we describe the framework developed to expose SiT
attacks. We evaluate the impact of SiT attack policies in Section
\ref{sec:eval} and we conclude the paper in Section \ref{sec:conc}
with a summary and future work.

\section{Related Work}\label{sec:relwork}
The work in this paper relates to two main areas of research: (1)
safety applications through V2V and V2I wireless communication and
(2) security in vehicular networks.\vspace{0.03in}

\noindent {\bf Traffic safety and management applications:} There
has been a large body of work in the area of ITS that utilizes
wireless signals for various safety and congestion management
applications. In \cite{misener2005cooperative}, the authors relied
on wireless communication to develop different cooperative
collision warning assistants for forward collision warning,
intersection collision and lane changes. In
\cite{Elbatt06cooperativecollision} the authors investigate the
impact of DSRC on the latency and the success probability in
Forward Collision Warning applications. The work in \cite{xu11}
proposes a safety application in which each vehicle is aware of
its nearest $k$ neighbors through V2V communication. The
architecture is envisioned for various safety scenarios, such as
collision avoidance, pre-crash sensing, traffic optimization and
lane changes warning. In \cite{dresner05}, Dresner et al. devised
a scheme in which vehicles can avoid congestion in intersections
by not stoping at all. The idea is that vehicles, through wireless
communications, reserve slots in space and time at the
intersection managers.\vspace{0.03in}


\noindent {\bf Security in vehicular networks:} There has also
been a large number of research studies that focused on the
security of vehicular networks. Leinm\"{u}ller et al. studied the
effects of false-position data on geographic routing in VANETs
\cite{leinmuller2005}. It was shown that malfunctioning and/or
malicious nodes broadcasting false position information can lead
to packet losses, routing delays and traffic interception, and
hence can drastically affect the performance, reliability and
security of position-based routing networks.\vspace{0.03in}

A model for attacks on inter-vehicle communication systems was
proposed in \cite{aijaz} wherein the goals and logistics of
various attacks are expressed in terms of attack trees. These
trees, which help understand and classify attacks, are used to
expose weaknesses and identify potential threats facing such
systems.\vspace{0.03in}

Stealth attacks whereby an attacker partitions an ad-hoc network
or hijacks traffic were studied in \cite{jakobsson:VTC03}. A key
idea is to keep a low exposure and to minimize the cost of the
attack through manipulation of the routing information of
well-behaving nodes.\vspace{0.03in}

Other attacks on vehicular networks include Sybil attacks used to
inject false messages into a vehicular network through use of
false identities \cite{Sybil2011}, DoS attacks through jamming the
communication channels, impersonation by using fake identities,
and bogus information attacks wherein wrong data could be diffused
in the network, for example to divert traffic from a given road.
In the SiT attack we consider herein, the propagation of false
information about traffic conditions is indirectly inflicted by
the attacker when the latter interferes with some of the signals
from the vehicles to their neighboring access points leading to
global traffic congestion. For a summary of various potential
attacks on vehicular networks we refer the reader to
\cite{raya2006} and references therein.

\section{A General Framework}\label{sec:model}
In this section we present a framework that enables the
identification of stealthy SiT attack policies that aim to create
traffic congestion.

\subsection{The Model}
\begin{figure}[h]
\centerline{\psfig{figure=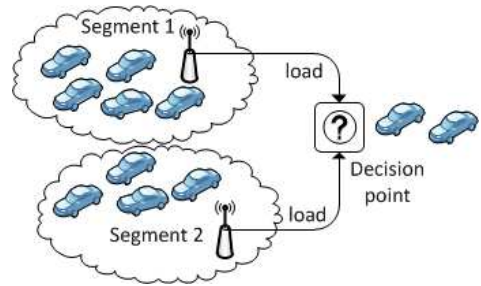, angle=270,width=2.6in}}\caption{A
vehicular network with 2 segments and a decision point.}
\label{fig:model}
\end{figure}

We consider a vehicular network that is composed of a set of
segments and a decision point. A segment is a portion of an
infrastructure (e.g., highway, bridge or a tunnel) that is
controlled by one access point. As vehicles utilize a segment,
they send wireless signals to the access point for the segment to
get an estimate of its current load (e.g., number of vehicles).
The access point reports its measurement back to a decision point
to influence future incoming traffic. Each segment presents an
alterative route to the driver. A decision point is a location at
which drivers must make an ``educated" decision on which segment
to use (e.g., at highway entrance points and intersections). At
each decision point, the load on each segment is presented to the
driver. Figure \ref{fig:model} shows a diagram describing the
setup.

We consider a discrete-time model in which at each time step, new
vehicles arrive at a decision point based on some arrival process.
For simplicity, we assume an infinite population of
vehicles for the arrival process. Based on the loads displayed, a
driver picks an appropriate segment. Vehicles exit each segment
based on its service rate.\vspace{0.03in}

Let $\lambda_k$ denote the arrival rate at the decision point at
time $k$, $\alpha_k(i), i = 1,\ldots ,n$, denotes the admission
ratio of vehicles on segment $i$ at time $k$, and $\beta_k(i),
i=1,\ldots, n$, denotes the service rate for segment $i$ at time
$k$, where $n$ is the total number of segments. Then the number of
vehicles, $q_k(i)$ on segment $i$ at time $k$ is given by:
\begin{eqnarray}
 \label{eqn:queue}
   q_k(i) &=& q_{k-1}(i) + \alpha_{k-1}(i)\lambda_k - \beta_k(i).
\end{eqnarray}
Throughout the paper, we assume that the service rates are known
and fixed.\vspace{0.03in}

Depending on the traffic optimization policies, the admission
ratio for each segment is determined based on the number of
vehicles on all segments according to the following equation:
\begin{eqnarray}
 \label{eqn:alpha}
   \alpha_k(i)= f \left(\hat q_{k-1}(1), \hat q_{k-1}(2),...\hat q_{k-1}(n)\right)
\end{eqnarray}
where $f$ is a traffic optimization function and $\hat q_k(i)$ is
the estimate of the queue length of segment $i$ at time $k$. For
example, the admission ratios can be chosen proportionally based
on the number of vehicles on each segment, weighted by the service
rate of each segment, or simply by picking the least loaded
segment.


\subsection{SiT Attacks}
The goal of SiT attacks is to cause traffic congestion by jamming
a subset of the wireless signals from the vehicles to the access
points leading to incorrect estimates displayed for drivers, and
consequently wrong decisions made by the drivers (e.g., choosing a congested segment).
To reflect their stealthy nature, an attacker pays a price
whenever he/she decides to jam a wireless signal. Clearly, if the
cost of jamming is very high, SiT attacks would not jam any signal
and if the cost of jamming is very low, SiT attacks would jam all
the time (i.e., DoS attack). We are interested in identifying
attacks policies that tradeoff damage and cost. In particular, we
are interested in identifying attack policies that can decide the
proper attack action based on the current state of the system.
\vspace{0.03in}

Let $s_k\in {\cal S}_k$ denote the state of the system at time
$k$, where ${\cal S}_k$ is the state space at time $k$. The state
of the system is the combination of the queue lengths $q_k(i)$ and
$\alpha_k(i), i = 1,\ldots, n$. Based on the number of new
arrivals and the admission ratios, the state is updated at the
next time step. 

The goal of the attacks is to unbalance the incoming traffic
across segments by selectively choosing what signal(s) to attack
at any state, if any. Let $u_k \in{\cal U}_k$ denote the control
action of the attacker at time $k$ and ${\cal U}_k$ the control
space at time $k$, which depends on the actual state $s_k$. Note that the estimates $\hat q_k(i), i = 1,
\ldots, n$, of the queue lengths are function of the attacker's
control action $u_k$ and the true queue lengths $q_k(i)$, i.e.,
\begin{eqnarray}
\hat q_k(i) = h(q_k(i), u_k) \label{eqn:estimate}
\end{eqnarray}
where $h$ is some function, which for simplicity is assumed to be
known to the attacker. Equations (\ref{eqn:queue}),
(\ref{eqn:alpha}) and (\ref{eqn:estimate}) define the state
dynamics. Note that from the attacker's standpoint, the state
$s_k$ consists not only of the queue lengths $q_k(i), i = 1,
\ldots, n$, but also the admission ratios $\alpha_{k}(i), i = 1,
\ldots, n$, since even for given values of the queue lengths, the
attacker's course of action will change depending on the
advertised admission ratios for the various segments.
\vspace{0.03in}

The attacker's action at time $k$ is obtained through a policy
$\mu_k$, which is defined as a mapping from the state space to the
control space, i.e., $\mu_k: {\cal S}_k\rightarrow{\cal
U}_k$.\vspace{0.03in}

Let $g(i,u,j)$ denote the reward obtained when the system evolves
from state $i$ to state $j$, under attack action $u$. The reward
can be described by the following equation:
\begin{eqnarray}
 \label{eqn:g}
   g(i,u,j) &=& d(i,u,j) - c(i,u,j)
\end{eqnarray}
where $d$ is the damage function and $c$ is the cost function of
the attack action $u$. An attack action becomes more appealing if
it can cause higher damage with less cost.\vspace{0.03in}

The infinite horizon expected reward is given by
\begin{eqnarray}
  \label{eqn:total_expected_reward}
 J(s_0, \mu_0, \mu_1, \ldots) = \textsf{E} \left[ \sum_{k=0}^{\infty} \gamma^k g(s_k, \mu_k(s_k), s_{k+1})|s_0\right]
\end{eqnarray}
where $s_0$ is the initial system state and $0<\gamma<1$ is a discount factor. Since the function $g(.)$
is bounded and $\gamma < 1$, the reward
function (\ref{eqn:total_expected_reward}) is well
defined.\vspace{0.03in}

The attacker is interested in maximizing the total expected discounted
rewards over the choice of attack policies. Hence, the goal is to
compute the solution to
\begin{eqnarray}
  \label{eqn:max}
 J^*(s_0) = \max_{\mu_0,\mu_1,\ldots} J(s_0, \mu_0, \mu_1, \ldots).
\end{eqnarray}
In other words, the adversary aims to maximize the cumulative
expected discounted reward over time by choosing
attack policies $\mu_0, \mu_1, \ldots$, as shown in
(\ref{eqn:max}). The problem now falls within the class of
infinite horizon problems with discounted reward. Hence, a stationary policy $\mu^*(.)$, i.e., which does not
depend on $k$, is optimal. The optimal attack policy can be
obtained by solving the following Bellman equation \cite{bellman}:
%
\begin{align}
  \label{eqn:j}
   J^*(s_0) = \max_{u\in{\cal U}(s_0)} \Bigg\{ &\textsf{E}[g(s_0, u, s_1 )] \nonumber\\
   &+ \gamma\sum_{s_1} p(s_1|s_0,u) J^*(s_1))\Bigg\}
\end{align}
\noindent where $J^*(.)$ is the optimal value function. The first
term on the R.H.S. represents the immediate stage reward in
(\ref{eqn:g}) and the second term is the future reward. The
conditional probability $p(.)$ is the probability of a transition
of the system to future state $s_{k+1}$ from state $s_k$ under
attack action $u$, and hence the summation in the second term is
over all possible future states, $s_1$ from $s_0$.
Solving the fixed point equation above gives the optimal tradeoff
between damage and cost from the standpoint of the
adversary.\vspace{0.03in}

Due to the large state space, solving the above equation may not
be computationally feasible. Thus, we propose an approximate
policy iteration method
\cite{getoor2007introduction,bertsekas2010approximate,sutton1998reinforcement}.
Before we describe the approximate policy iteration methodology,
we provide some brief background on exact policy
iteration.\vspace{0.03in}

Exact Policy iteration consists of 2 steps: policy evaluation and
policy improvement. In the policy evaluation step, we start with
an initial policy $\mu$. Then, we solve a system of linear
equations to evaluate the cost function $J_{\mu}(s)$ starting from
state $s$ and using policy $\mu$:

\begin{eqnarray}
  \label{eqn:pol_eval}
   J_{\mu}(s) &=& \sum_{s'} p(s'|s,\mu(s)) \left(g(s,\mu(s), s') + \gamma J_{\mu}(s')\right)
\end{eqnarray}

\noindent where the summation is over the set of states $s'$ that
can be reached from state $s$ and $g(s,\mu(s),s')$ is the reward
obtained from the transition from $s$ to every state in $s'$ under
policy $\mu(s)$. In the policy improvement step, an improved
policy $\bar \mu$ is generated according to the following
equation:

\begin{eqnarray}
  \label{eqn:pol_imp}
   \bar\mu(s) &=& \arg\max_{u \in {\cal U}(s)} \sum_{s'} p(s'|s,u) \left(g(s,u,s') + \gamma J_{\mu}(s')\right).
\end{eqnarray}

\noindent The improved policy is the one that maximizes the reward
through selecting the best attack action $u$, from the set of
actions ${\cal U}(s)$ available from state $s$. The improved
policy $\bar \mu$ is then used as the new policy and a new
iteration starts.\vspace{0.03in}

\noindent One of the main challenges with exact policy iteration
is the size of the state space. Even for a small system with 2
segments with each potentially holding up to 100 vehicles, the
size of the state space is $100\times 100$, without accounting for
the $\alpha$ which has a similar order of
magnitude.\vspace{0.03in}


In the approximate policy iteration variant, we run Monte Carlo
simulations to evaluate the current policy rather than solving the
system of linear equations. We approximate $J_{\mu}(s)$ with a
parametric representation $\tilde J_r(s)$:

\begin{eqnarray}
  \label{eqn:pol_imp_approx}
    \tilde J_r(s) &=& \sum_{j=1}^{M} r_j\phi_j(s)
\end{eqnarray}

\noindent where $\phi$ is a column of features, $r$ is a row of
weights (one for each feature), and $M$ is the number of those
features. The idea is to extract $M$ features that characterize
state $s$ and approximate $J_{\mu}(s)$ by selecting $r$ that
solves a least square problem between the rewards obtained from
the Monte Carlo simulations and the cross product of
$r_j\phi_j(s)$. It is known that the linear combination of well
chosen features can capture essential nonlinearities in the reward
function \cite{busoniu2011cross,di2010adaptive,yu2009basis}.

\subsection{Feature Selection}
Due to the approximate nature of our proposed methodology, we must
rely on a set of representative features to capture the
fundamental characteristics of the state. We used the following
features to approximate the value function $\tilde J_r(s)$ for
state $s$ with:

\begin{enumerate}
    \item The number of vehicles on each segment.
    \item The degree of imbalance between the number of vehicles on
    each segment (weighted by their departure rates when segments have different service rates).
    \item The segment that is the least occupied.
    \item The admission ratios reported to the drivers.
    \item Difference between the true admission ratio (that is based on the true occupancy of vehicles) and the one
    utilized by drivers at the decision point.
    \item How far the admission ratio is from the ideal one (e.g., 0.5 in case of two identical
    segments).
    \item A flag to indicate whether the two segments are empty or
    not.
\end{enumerate}

\subsection{Performance Metrics}
Throughout this work, we focus on different metrics/functions for
assessing damage and cost. The damage function due to a SiT attack
can be instantiated as the imbalance between different segments,
weighted by their service rates. Another possible instantiation is
the gap between the admission ratios reported to the drivers and
the ``true" admission ratios that {\it should have been reported}.
Our model can also be easily extended to account for other forms
of damage such as queueing delays and other factors that typically
occur under congestion. For example, it is expected that the
service rates of segments decrease as the number of vehicles
increase. Similarly, different instantiation for the cost function
are possible. The motivation behind the ones we consider in this
paper is that as the number of attacked vehicles increases, the
attack can be more exposed and thus the cost of the attack should
reflect such a greater risk of exposure.\vspace{0.03in}

Since the attacker must select an attack cost, it is not clear
what would be a reasonable choice for a given problem. By varying
the attack cost, however, an attacker can discover different
polices. In this paper, we are interested in stealthy attacks,
thus we focus on a smaller region of the attack cost values where
the resulting policy is not a complete DoS-like attack nor a
no-attack policy. This is obtained through trial and error until
the interesting region is found.

\section{Performance Evaluation}\label{sec:eval}
In this section we report on our evaluation of the approximate
policy iteration (API) methods on a number of systems that are
instantiated from the model described in Section \ref{sec:model}.
In this paper we limit our evaluation to systems with two
alternate segments. We fix the discount factor $\gamma$ to 0.99.

\subsection{System One}
Consider a system composed of 2 segments and 1 decision point.
Vehicles arrive to the decision point per unit time based on the
following probability distribution: 3 vehicles with probability
0.3, 8 vehicles with probability 0.6 and 30 vehicles with
probability 0.1. Thus the average arrival rate is 8.7. We assume
the two segments are identical and each one has a maximum service
rate of 5 vehicles per unit time. Based on the reported number of
vehicles on each segment, the decision point reflects the
admission ratio for each segment to balance traffic between the
segments. In this system we assume that all the drivers follow the
information displayed.\vspace{0.03in}

An attacker mounting a SiT attack jams a subset of the signals
from vehicles to the access point. We assume that a SiT attack
only affects 50\% of the vehicles. Thus, the estimate $ \hat
q_k(i)$ in (\ref{eqn:estimate}) becomes $ \hat q_k(i) =
\frac{1}{2} \times q_k(i)$ whenever the attacker decides to
attack. We take the cost function, $c$, to be a constant value,
$C_T$, multiplied by the number of vehicles
affected.\vspace{0.03in}

In this system, the attacker aims to unbalance the traffic between
the two segments. We instantiate the damage function, $d$, to be
the absolute difference between the number of vehicles on each
segment:

\begin{eqnarray}
 \label{eqn:exp_damage}
   d &=&  |q_k(1) - q_k(2)|.
\end{eqnarray}

At any state, the attacker can choose between the following
actions:

 \begin{enumerate}
    \item Not attack with cost 0.
    \item Attack half the vehicles on segment 1 with cost $C_T \times 0.5 \times q_k(1)$.
    \item Attack half the vehicles on segment 2 with cost $C_T \times 0.5 \times q_k(2)$.
 \end{enumerate}

We start our approximate policy iteration method from 32
representative states that are chosen at 25 increments to give
good coverage of the state space. Moreover, half those states
reflect the true admission ratio while the other half have
admission ratios chosen at random. We start with a random policy
as a roll-out one. From each representative state, we run 50
independent trajectories and we compute the average reward across
them. In each trajectory, we simulate the attack policy for 100
steps. In each iteration, a new policy is generated and we keep
track of the weight vector $r$ that produces the policy with the
maximum reward.\vspace{0.03in}

Once the weight vector $r$ is obtained, we compare between
policies based on a completely different set of states that are
generated at random. In other words, there is no intentional
overlap between our training data and the ones we use for
evaluation.\vspace{0.03in}

It is important to note that with approximate policy iteration
methods, there are no guarantees that the system will converge
(i.e., no guarantee that the resulting policy is an improvement of
the previous one) as with exact policy iteration. Thus, we do not
have a termination method except to run for a large number of
iterations and to choose the best policy. We typically use between
100 and 1000 iterations.



\begin{figure}[ht]
\centerline{\psfig{figure=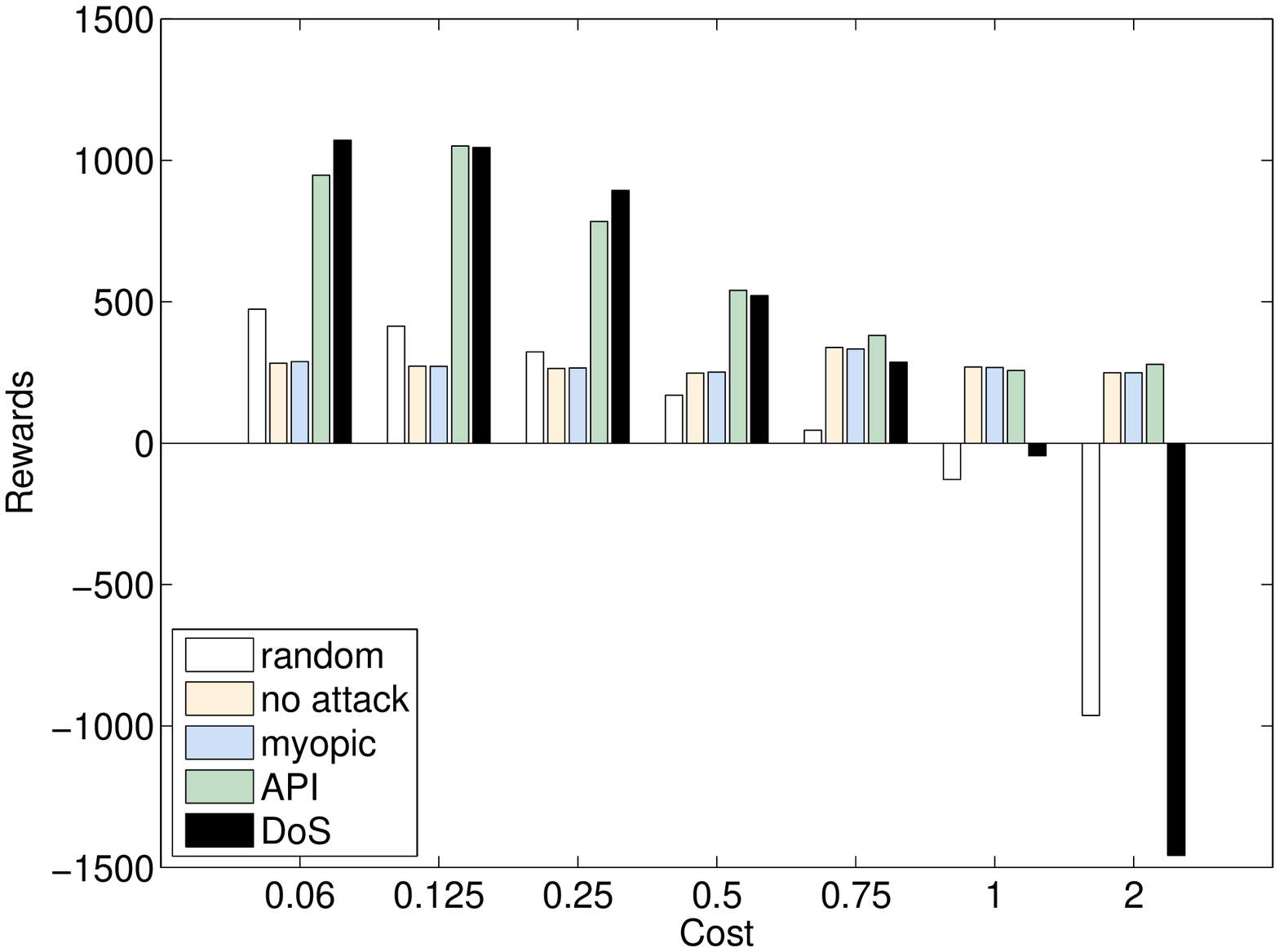,
angle=0,width=0.41\textwidth}}\centerline{\psfig{figure=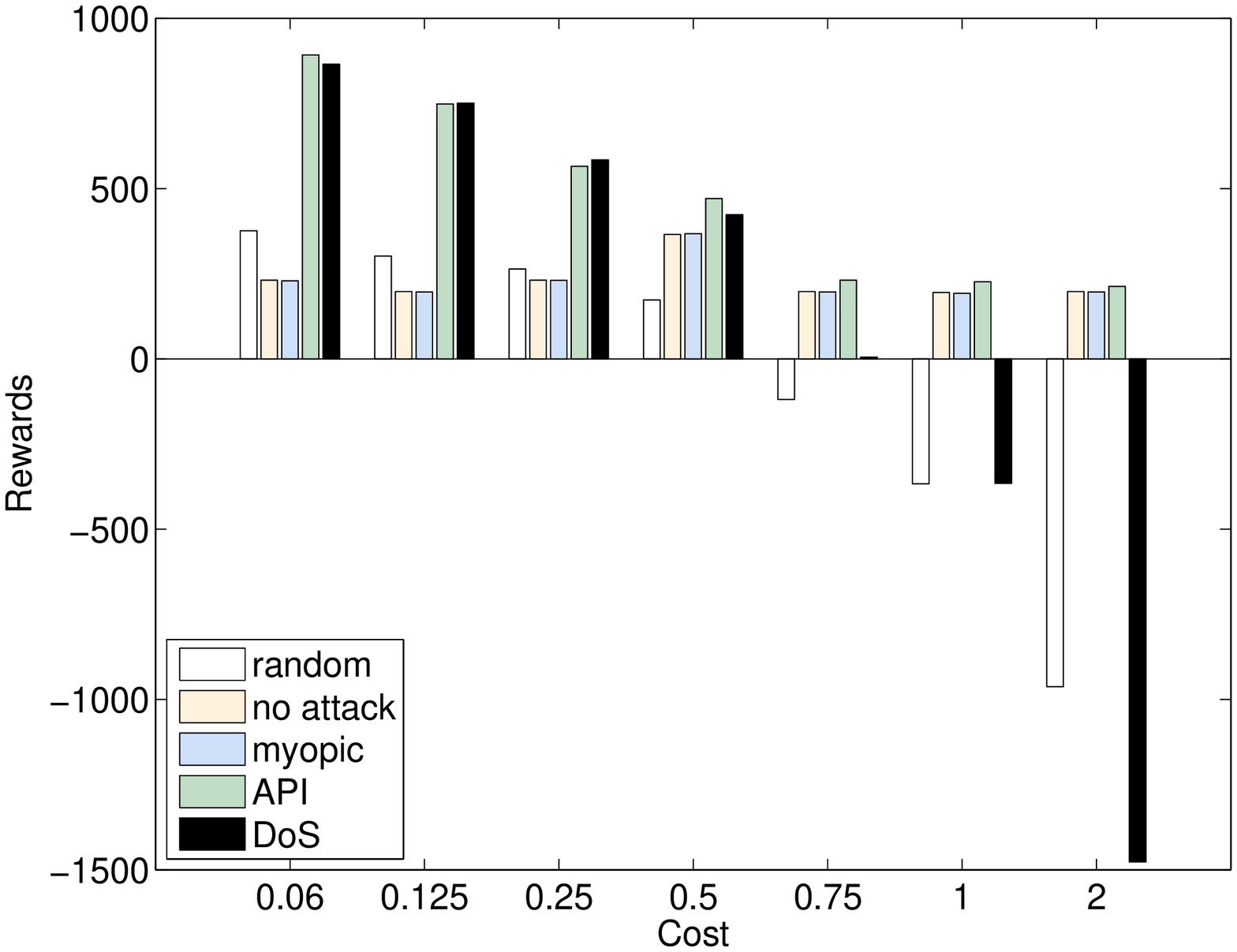,
angle=0,width=0.41\textwidth}}\centerline{\psfig{figure=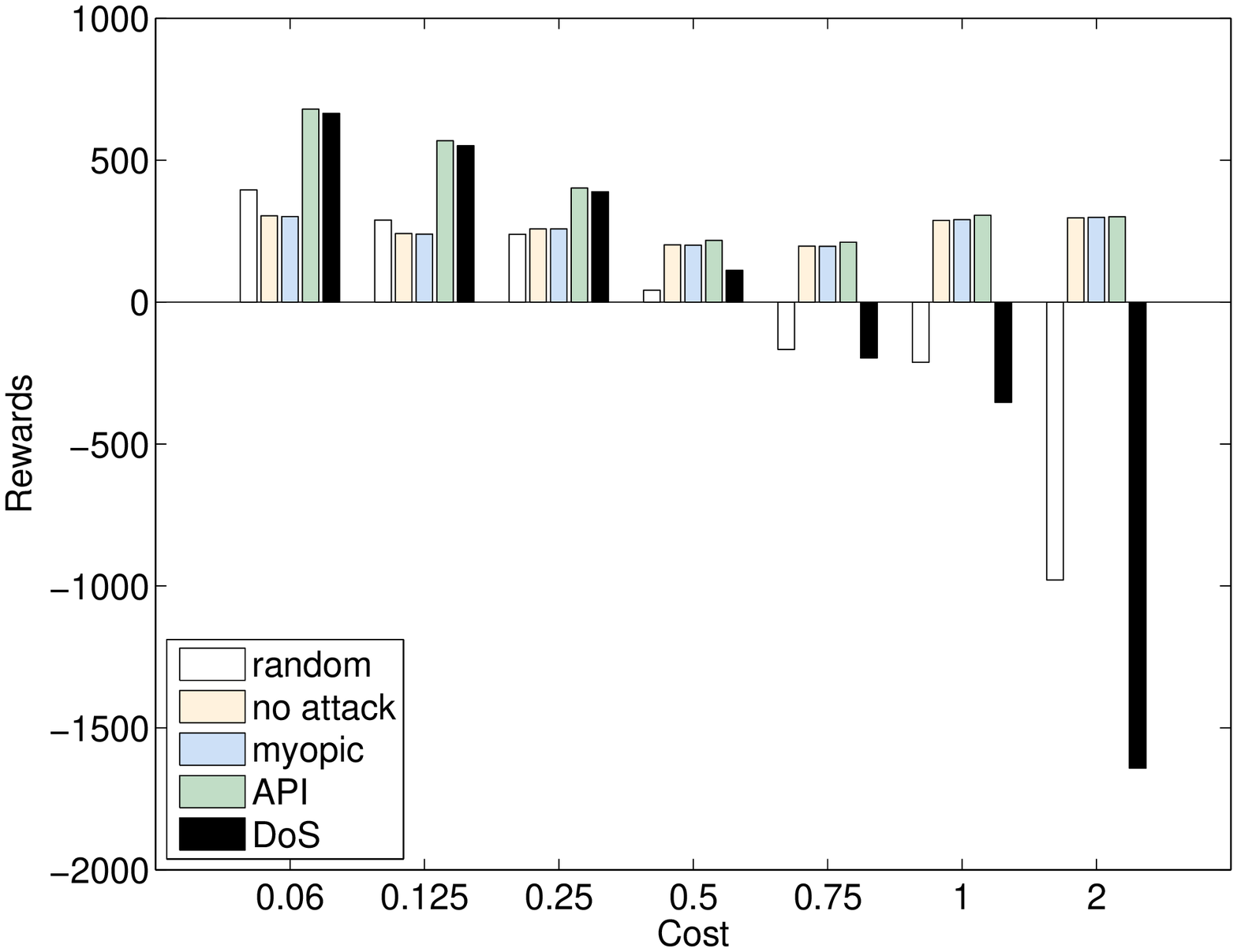,
angle=0,width=0.41\textwidth}}\caption{Comparison between API,
myopic, random, DoS and no attack for System One under different
attack costs. Attack success rate is 100\% (top), 75\% (center)
and 50\% (bottom).} \label{fig:system1}
\end{figure}

Figure \ref{fig:system1} shows the rewards obtained for different
cost values $C_T$ under different policies. Figure
\ref{fig:system1} (top) is for attack success rate 100\%, Figure
\ref{fig:system1} (center) is for attack success rate 75\%, and
Figure \ref{fig:system1} (bottom) is for attack success rate 50\%.
We compare our Approximate Policy Iteration (API) to a no-attack
policy, a random attack, a DoS attack on one of the segments, and
a myopic attack (in which only the immediate reward is used to
select an action without regard to the future reward). We only
show the interesting region based on the attack costs. If the cost
of the attack is very low, API matches a DoS attack and if the
cost of the attack is very high, API matches a no-attack policy.
One can see that API tracks the best policies very well and in the
majority of the cases it provides the policy with the highest
reward.\vspace{0.03in}

Notice also that the performance of the API method appears to
improve as the degree of certainty in the attack success rate
decreases. With an attack success rate of 50\%, the API method was
consistently better than all policies across all costs, whereas
with higher attack success rates, the performance may be slightly
less than some policies. This is the case because the API method
takes the success rate into account which choosing actions that
achieve a balance between the immediate and the future
rewards.\vspace{0.03in}

\begin{figure}[h]
\centerline{\psfig{figure=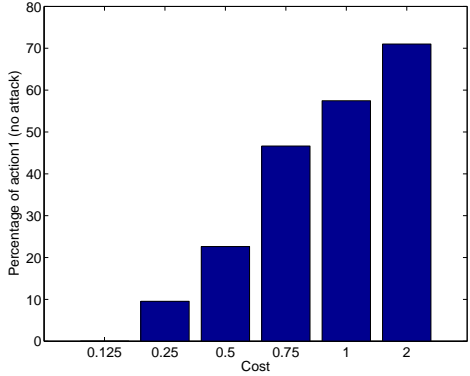,angle=0,width=0.4\textwidth}}\caption{Percentage
of no-attack actions as we vary the cost of the attack. Results
are for System one with 75\% success rate.}
\label{fig:system1actions}
\end{figure}

Figure \ref{fig:system1actions} shows the percentage of time the
no-attack action was selected as we vary the cost of the attack.
The results are shown for System one with 75\% attack success
rate. One can see that our proposed policy smoothly adjusts the
level of aggression based on the cost of the attack.





\begin{table}[ht]
\centering
\begin{tabular}{|c|c|c|c|c|c|c|c|}
  \hline
  Attack Cost & 0.06 & 0.125 & 0.25 & 0.5 & 0.75 & 1 & 2\\
  \hline
  50\% (500 iter) & 473 & 357 & 238 & 248 & 318 & 378 & 375\\
    \hline
  75\% (100 iter) & 98 & 32 & 2 & 79 & 76 & 65 & 62\\
  \hline
  100\% (1000 iter) & 812 & 970 & 659 & 328 & 460 & 284 & 669\\
  \hline
\end{tabular}\caption{Number of iterations to find the best policy for system one under different success probabilities (50\%, 75\% and 100\%).}\label{tbl:system1_conv}
\end{table}

Table \ref{tbl:system1_conv} shows the number of iterations it
took to reach the best policy for System One under different
attack costs and for different attack success probabilities.
Throughout our evaluation, we limited the number of iterations
below 1000. Finding the best policy changes from a system to
another and depends on our choice of roll-out policies and on the
randomization within the framework. We listed those values here as
a mean to share our experience with the API methods.

\subsection{System Two}
Our second system is a variant of the first system, but covers a
wider attack scope. In particular, we assume that an attacker can
decide between different SiT attack intensities by choosing
between the following actions:

\begin{enumerate}
    \item Not attack with cost 0
    \item Attack 25\% of vehicles on segment 1 with cost $C_T \times 0.25 \times q_k(1)$
    \item Attack 50\% of vehicles on segment 1 with cost $C_T \times 0.5 \times q_k(1)$
    \item Attack 75\% of vehicles on segment 1 with cost $C_T \times 0.75 \times q_k(1)$
    \item Attack 25\% of vehicles on segment 2 with cost $C_T \times 0.25 \times q_k(2)$
    \item Attack 50\% of vehicles on segment 2 with cost $C_T \times 0.5 \times q_k(2)$
    \item Attack 75\% of vehicles on segment 2 with cost $C_T \times 0.75 \times q_k(2)$.
\end{enumerate}

\begin{figure*}[ht]
\centerline{\psfig{figure=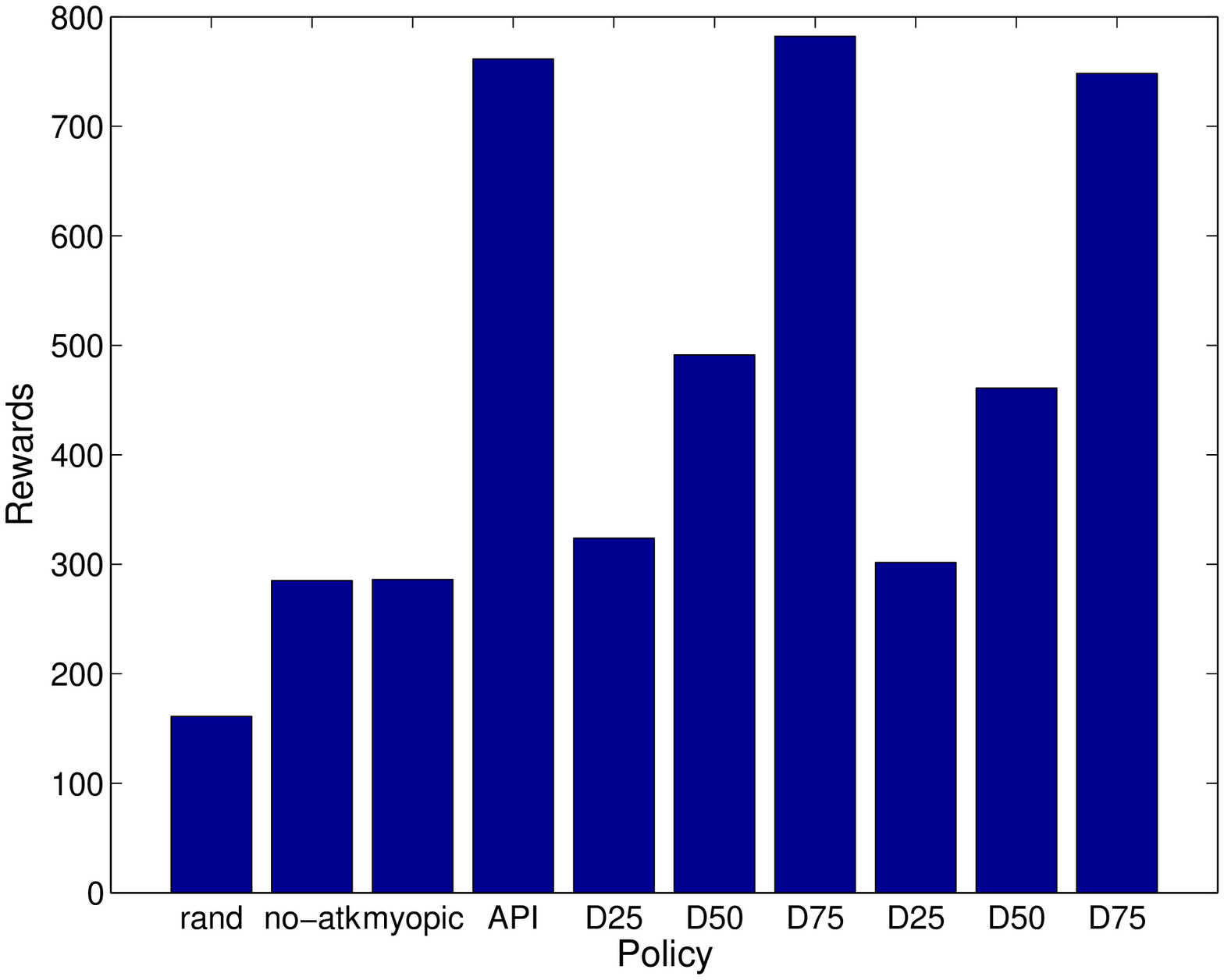,
angle=0,width=0.32\textwidth}\psfig{figure=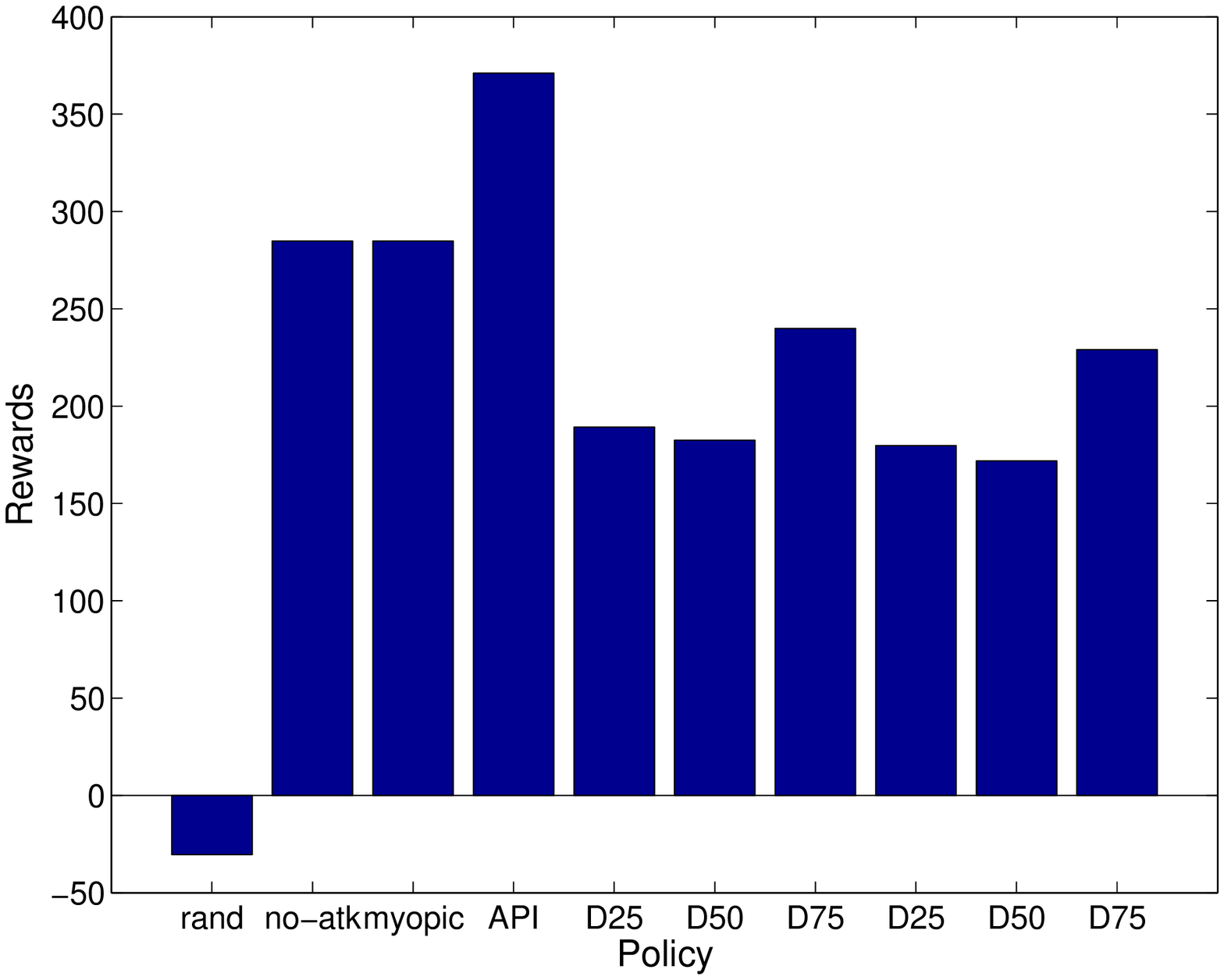,
angle=0,width=0.32\textwidth}\psfig{figure=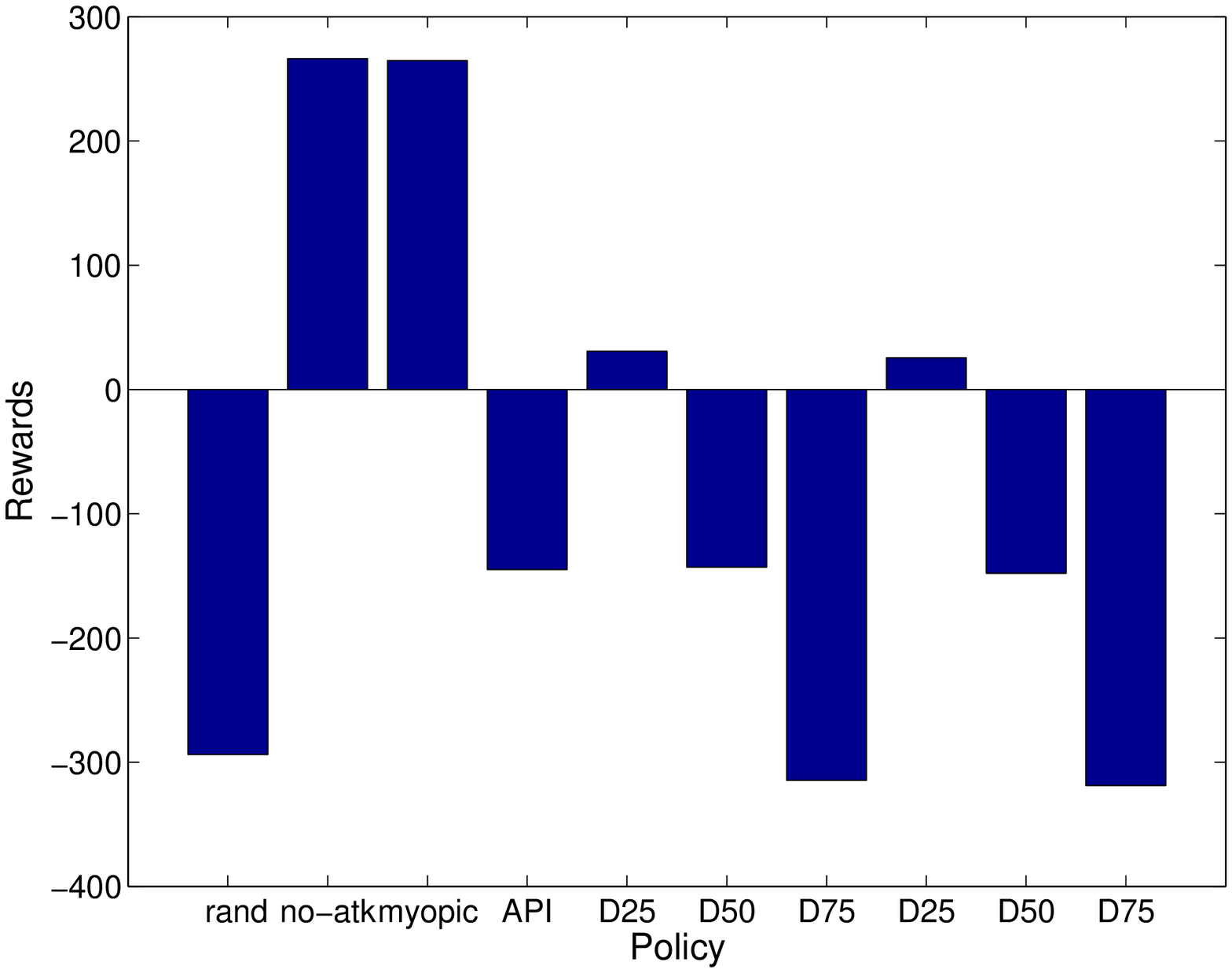,
angle=0,width=0.32\textwidth} }\caption{System 2 with 6 actions.
Cost 0.5 (left), cost 0.75 (center). and cost 1 (Right). We
compare between random, no attack, myopic, API, and DoS attacks at
various levels. D25, D50 and D75 indicate attacking 25\%, 50\% and
75\% of the vehicles on a segment, respectively (results are shown
for segment one first and then for segment two).}
\label{fig:system2}
\end{figure*}

Figure \ref{fig:system2} shows results for System Two under three
different attack costs (0.5, 0.75 and 1). We compare between
random, no attack, myopic, API, and DoS attacks at various levels
on both segments (D25, D50 and D75 indicate attacking 25\%, 50\%
and 75\% of the vehicles on a segment, respectively). When the
cost of the attack is low (left), API matches the most aggressive
DoS attack. When the cost is 0.75 (center), API outperforms all
the policies. Note, however, that under attack cost, 1, we failed
to find a good policy, since API does worse than some of other
policies. As noted above, this can happen with approximation
methods. The algorithm can get stuck in local maxima or needs more
iterations to find better policies. One approach to tackle this
issue is to force different roll-out policies or introduce
randomization within the obtained policy (e.g., choose a subset of
actions at random regardless of the policy)
\cite{bertsekas-neuro}. We leave this issue for future
investigation.

\begin{figure*}[thb]
\centering
\begin{minipage}{0.42\textwidth}
\centerline{\psfig{figure=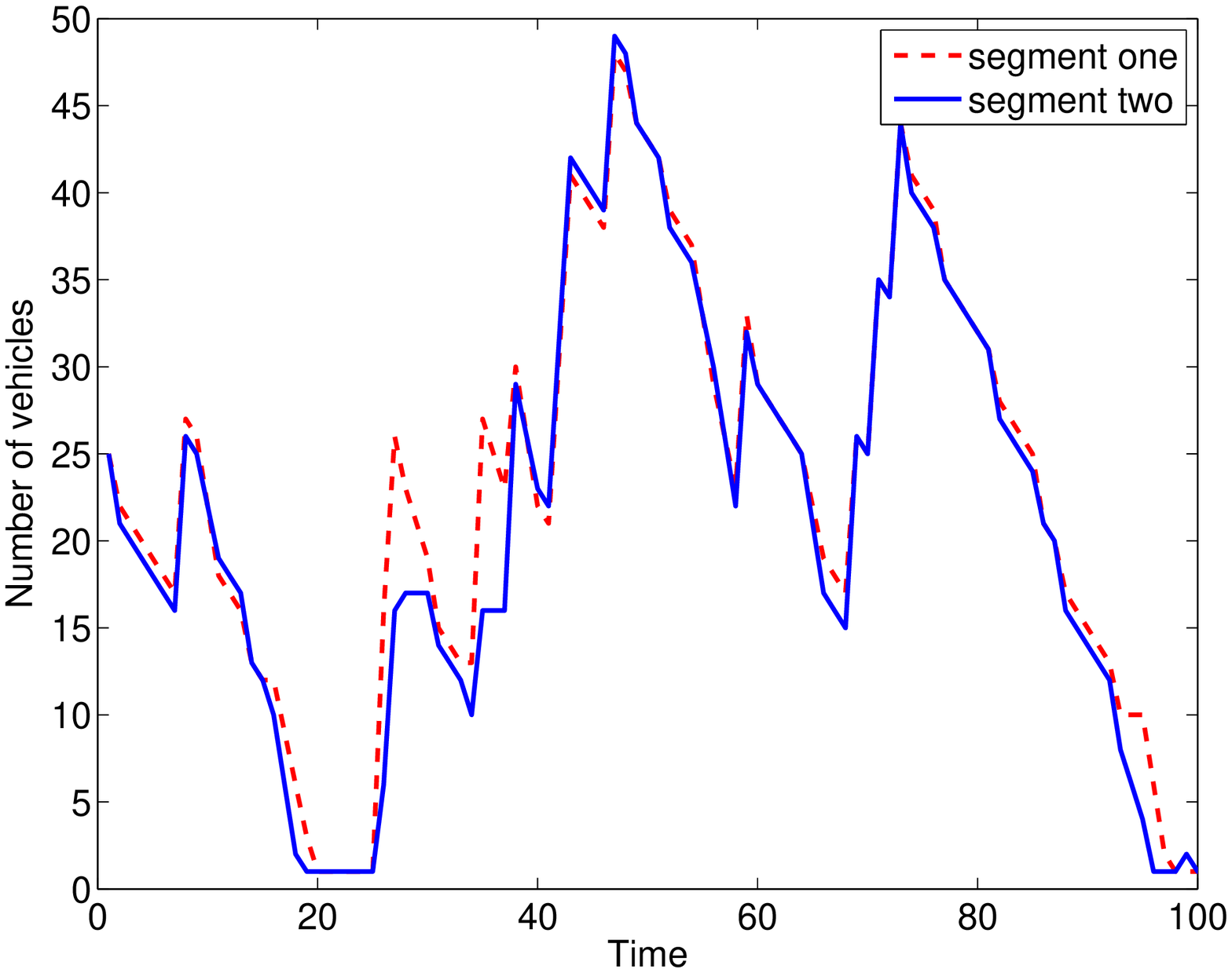,angle=0,width=0.97\textwidth}}
\centerline{(a)}
\end{minipage}
\begin{minipage}{0.42\textwidth}
\centerline{\psfig{figure=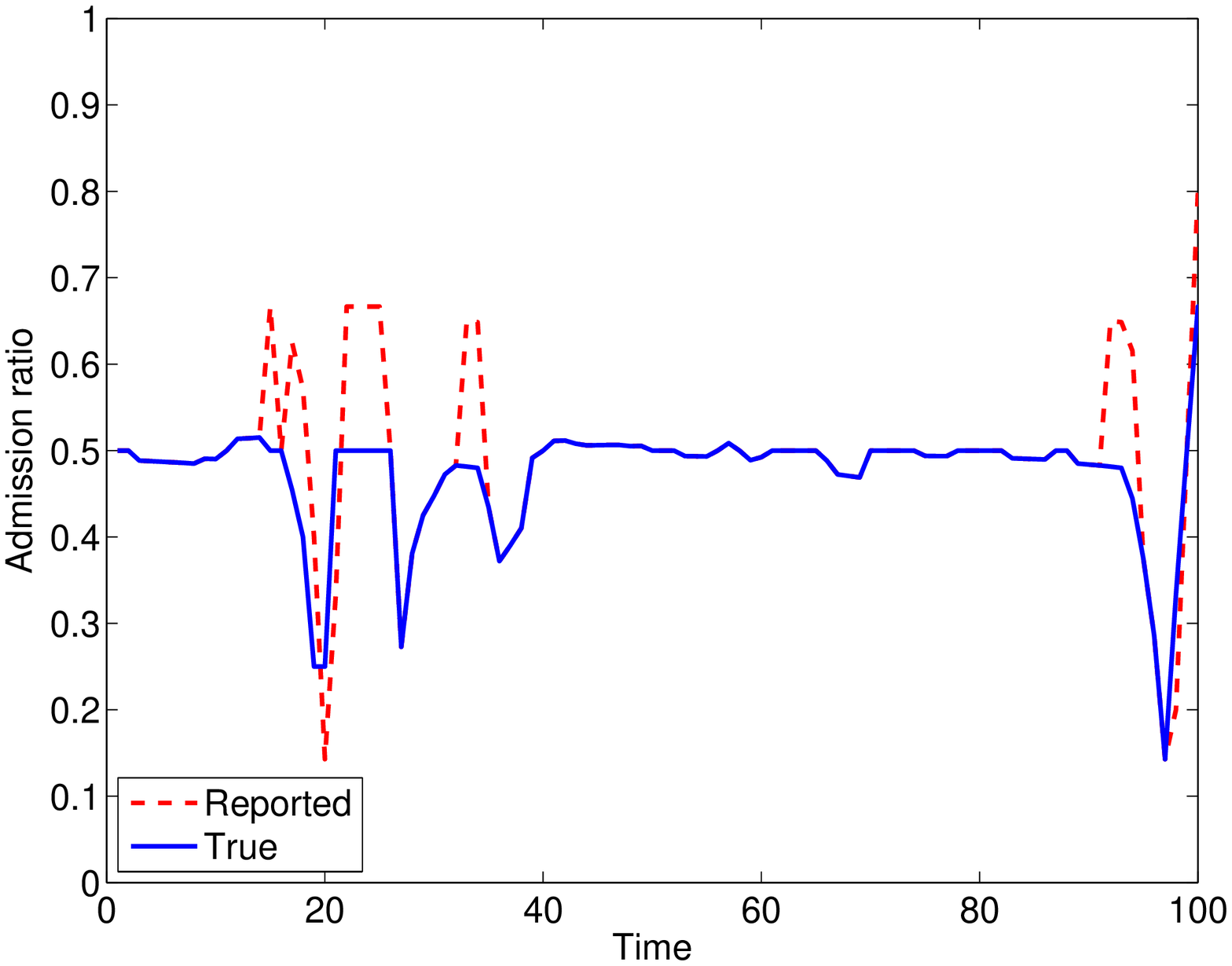,angle=0,width=0.97\textwidth}}
\centerline{(b)}
\end{minipage}
\begin{minipage}{0.42\textwidth}
\centerline{\psfig{figure=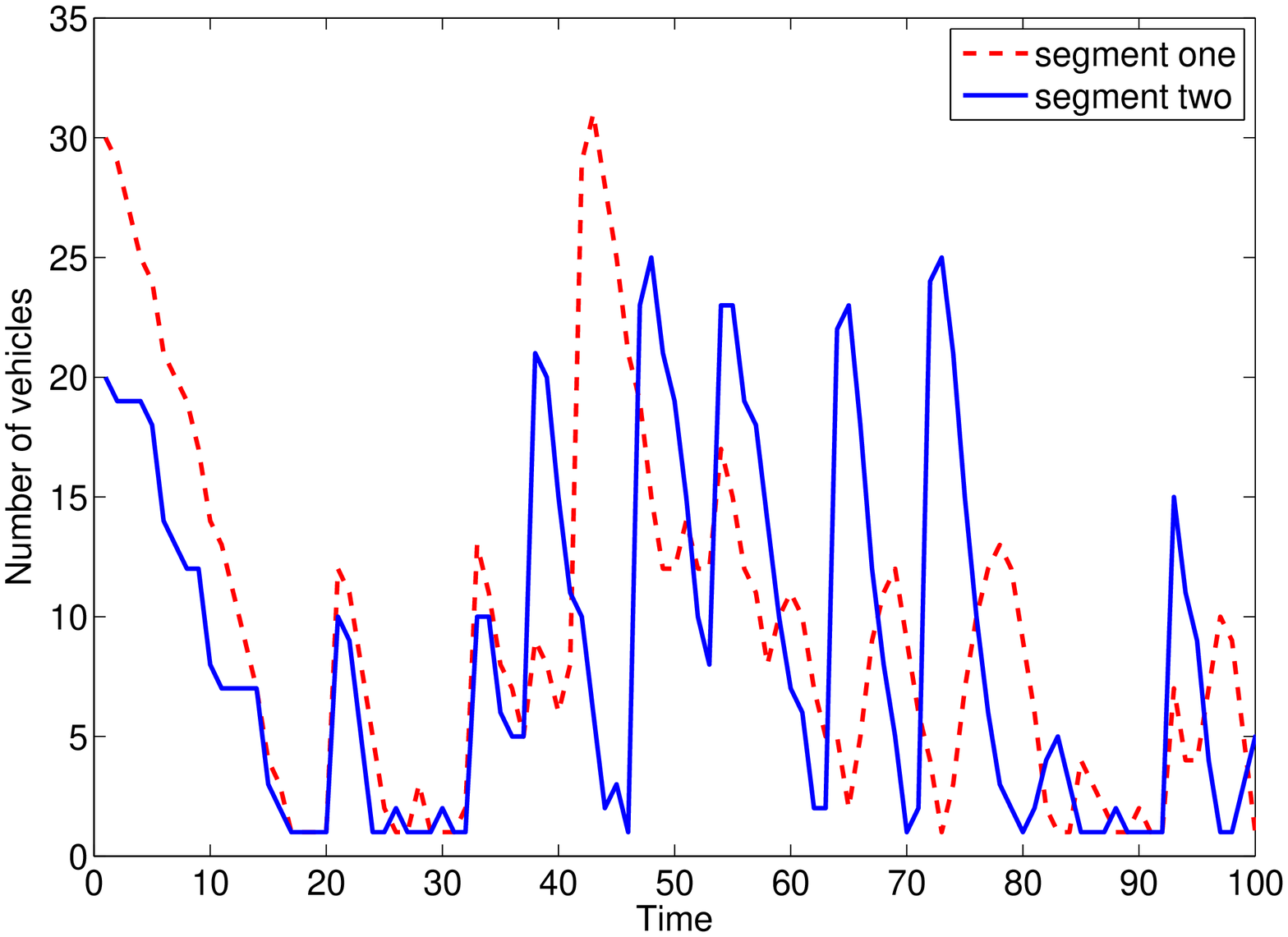,angle=0,width=0.97\textwidth}}
\centerline{(c)}
\end{minipage}
\begin{minipage}{0.42\textwidth}
\centerline{\psfig{figure=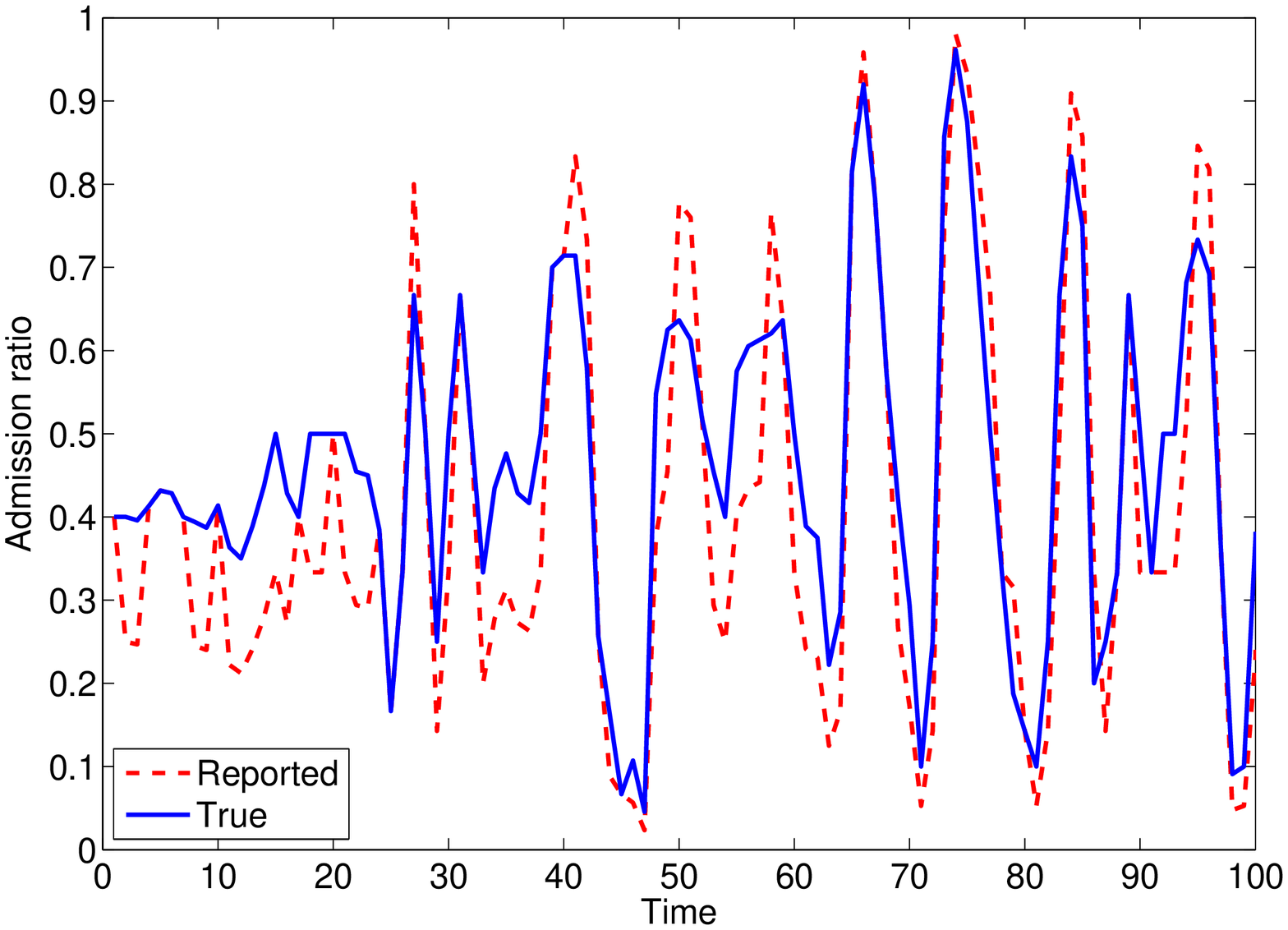,angle=0,width=0.97\textwidth}}
\centerline{(d)}
\end{minipage}

\caption{Comparison between identical service rate segments (top
row) and different service rate ones (bottom row) under the same
attack cost. (a) and (c) show the number of vehicles on each
segment. (b) and (d) show the admission ratio comparing the one
reported to the driver versus the true admission ratio.}
\label{fig:system3}
\end{figure*}

\subsection{System Three}
In this system, we consider the case when segments have different
service rates. To make valid comparisons, we use the exact system
as System One except that the maximum service rate for segment 1
is 4 vehicles per unit time and for segment 2 is 6 vehicles per
unit time. Both systems started with the same total number of
vehicles weighted by their service rates. The initial admission
ratio reflected the true admission ratio. Figure \ref{fig:system3}
shows the number vehicles on each segment and the admission ratio
for System One (top row) and those for System Three (bottom row).
Results are presented for Cost 2. When the attacker decides not to
attack, the admission ratio reported matches the true admission
ratio.

\begin{figure}[h]
\centerline{\psfig{figure=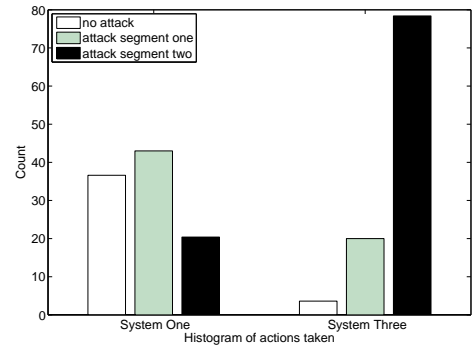,
angle=0,width=0.4\textwidth}}\caption{Histogram of the actions
taken under the same attack cost.} \label{fig:system1_3}
\end{figure}

We have found that, under the same cost, attacks on systems that
have segments with different service rates lead to higher attack
reward than attacks on systems that have segments with identical
service rates. Figure \ref{fig:system1_3} shows the average number
of different actions taken by the attacker under cost 2, for both
systems. One can see that for system three, the best attack policy
attacks either one of the segments more than 90\% of the time when
compared to system one in which the best attack policy attacks
either one the segments only around 65\% of the time.

\section{Conclusions}\label{sec:conc}
Traffic safety applications are increasingly relying on wireless
technologies in transforming our transportation system through
empowering drivers to make good decisions, improving their safety
and reducing the overall cost. Currently, the development of many
ITS applications and standards are underway. Thus, it is crucial
to expose vulnerabilities at this early phase before deployment
and to have the appropriate defense mechanisms in place once these
systems become operational. In this paper, we developed a
framework that is capable of exposing stealthy SiT attacks that
aim to cause traffic congestion by selectively interfering with a
subset of the signals from vehicles to the infrastructure. We have
evaluated the generated attack policies and demonstrated their
potency when compared to other policies such as myopic, random and
DoS attacks. Unlike other policies, the proposed policy
judiciously adapts to the system parameters (e.g., queue lengths,
costs, and service rates) to select attack actions that balance
between the current stage and future rewards. Moreover, we have
shown that our proposed policy performs better as the degree of
uncertainty in the system increases, making it appealing to
adversaries that may not be confident of the exact impact of the
attack. Furthermore, through our evaluation we have demonstrated
that systems that employ segments with different service rates are
more susceptible to attacks than those employing segments with
similar service rates. To the best of our knowledge, this work is
the first to look at the effect of jamming attacks through an MDP
framework and to apply approximation techniques to identify
optimal/suboptimal stealthy policies.\vspace{0.03in}

We are currently investigating the impact of feature selection on
the resulting policies and what constitutes a good set. We are
also looking at systems in which the service rates for the
different segments changes based on the number of vehicles
present. Another direction we are working on is the development of
defense techniques against SiT attacks. In particular, we are
looking at the use of randomization techniques to prevent an
attacker from implicitly adjusting the admission ratios.


\section*{Acknowledgment}
This work is supported by the CNS NSF grant \#1149397.

\bibliographystyle{IEEE}
\bibliography{mina}

\begin{thebibliography}{10}

\bibitem{Elbatt06cooperativecollision}
T.~Elbatt, S.~Goel, G.~Holl, H.~Krishnan, and J.~Parikh,
\newblock ``{Cooperative Collision Warning Using Dedicated Short Range Wireless
  Communications},''
\newblock in {\em Proceedings of the 3rd international workshop on Vehicular
  Ad-hoc Networks (VANET)}, Los Angeles, CA, September 2006.

\bibitem{misener2005cooperative}
J.~Misener, R.~Sengupta, and H.~Krishnan,
\newblock ``{Cooperative Collision Warning: Enabling Crash Avoidance with
  Wireless Technology},''
\newblock in {\em Proceedings of 12th World Congress on ITS}, San Francisco,
  CA, 2005.

\bibitem{yan2010cooperative}
G.~Yan, W.~Yang, M.C. Weigle, S.~Olariu, and D.~Rawat,
\newblock ``{Cooperative Collision Warning through Mobility and Probability
  Prediction},''
\newblock in {\em IEEE Intelligent Vehicles Symposium (IV)}, San Diego, CA,
  June 2010.

\bibitem{tatchikou2005cooperative}
R.~Tatchikou, S.~Biswas, and F.~Dion,
\newblock ``{Cooperative Vehicle Collision Avoidance using Inter-vehicle Packet
  Forwarding},''
\newblock in {\em Proceedings of IEEE Globecom}, St. Louis, MO, December 2005.

\bibitem{biswas2006vehicle}
S.~Biswas, R.~Tatchikou, and F.~Dion,
\newblock ``{Vehicle-to-Vehicle Wireless Communication Protocols for Enhancing
  Highway Traffic Safety},''
\newblock {\em IEEE Communications Magazine}, vol. 44, no. 1, pp. 74--82, 2006.

\bibitem{FCC}
Federal~Communications Commission,
\newblock ``{Dedicated Short Range Communications (DSRC) Service},''
  \url{http://wireless.fcc.gov/services/index.htm?job=about&id=dedicated_src}.

\bibitem{luosurvey}
J.~Luo and J.P. Hubaux,
\newblock ``{A survey of Research in Inter-Vehicle Communications},''
\newblock {\em Embedded Security in Cars-Securing Current and Future Automotive
  IT Applications}, pp. 111--122, 2006.

\bibitem{sichitiu2008inter}
M.~Sichitiu and M.~Kihl,
\newblock ``{Inter-vehicle Communication Systems: A Survey},''
\newblock {\em IEEE Communications Surveys \& Tutorials}, vol. 10, no. 2, pp.
  88--105, 2008.
\newpage
\bibitem{jamming}
W.~Xu, K.~Ma, T.~Wade, and Y.~Zhang,
\newblock ``{Jamming Sensor Networks: Attacks and Defense Strategies},''
\newblock {\em IEEE Network}, 2006.

\bibitem{thuente2006intelligent}
D.~Thuente and M.~Acharya,
\newblock ``{Intelligent Jamming in Wireless Networks with Applications to
  802.11b and other Networks},''
\newblock in {\em Proceedings of IEEE MILCOM}, Washington, DC, October 2006.

\bibitem{sang2009capabilities}
L.~Sang and A.~Arora,
\newblock ``{Capabilities of Low-power Wireless Jammers},''
\newblock in {\em Proceedings of INFOCOM}, Rio de Janeiro, Brazil, April 2009.

\bibitem{reactivejamming2011}
M.~Wilhelm, I.~Martinovic, J.~Schmitt, and V.~Lenders,
\newblock ``{Short Paper: Reactive Jamming in Wireless Networks: How Realistic
  is the Threat?},''
\newblock in {\em Proceedings of the fourth ACM conference on Wireless network
  security}, 2011.

\bibitem{pelechrinis2011denial}
K.~Pelechrinis, M.~Iliofotou, and V.~Krishnamurthy,
\newblock ``{Denial of Service Attacks in Wireless Networks: The Case of
  Jammers},''
\newblock {\em IEEE Communications Surveys \& Tutorials}, vol. 13, no. 2, 2011.

\bibitem{law2009energy}
Y.~Law, M.~Palaniswami, L.~Hoesel, J.~Doumen, P.~Hartel, and P.~Havinga,
\newblock ``{Energy-Efficient Link-Layer Jamming Attacks gainst Wireless Sensor
  Network MAC Protocols},''
\newblock {\em ACM Transactions on Sensor Networks (TOSN)}, vol. 5, no. 1, pp.
  1--38, 2009.

\bibitem{diymedia}
Diymedia,
\newblock ``{FCC Enforcement News},''
  \url{http://www.diymedia.net/fccwatch/enfnews.htm}.

\bibitem{raya}
M.~Raya and J.~Hubaux,
\newblock ``{Securing Vehicular Ad Hoc Networks},''
\newblock {\em Journal of Computer Security}, vol. 15, no. 1, pp. 39--68, 2007.

\bibitem{aijaz}
A.~Aijaz, B.~Bochow, F.~D{\"o}tzer, A.~Festag, M.~Gerlach, R.~Kroh, and
  T.~Leinm{\"u}ller,
\newblock ``{Attacks on Inter Vehicle Communication Systems- An Analysis},''
\newblock in {\em Proceedings of the International Workshop on Intelligent
  Transportation (WIT)}, Hamburg, Germany, March 2006.

\bibitem{mobilityreport}
Texas~Transportation Institute,
\newblock ``2011 urban mobility report,''
  \url{http://tti.tamu.edu/documents/mobility-report-2011.pdf}.

\bibitem{xu11}
O.~Wolfson B.~Xu and H.~Ju Cho,
\newblock ``Monitoring neighboring vehicles for safety via v2v communication,''
\newblock in {\em IEEE International Conference on Vehicular Electronics and
  Safety (ICVES)}, 2011.

\bibitem{dresner05}
K.~Dresner and P.~Stone,
\newblock ``Multiagent traffic management: An improved intersection control
  mechanism,''
\newblock in {\em In The Fourth International Joint Conference on Autonomous
  Agents and Multiagent Systems}, Utrecht, The Netherlands, July 2005.

\bibitem{leinmuller2005}
T.~Leinm\"{u}ller, E.~Schoch, F.~Kargl, and C.~Maih\"{o}fer,
\newblock ``{Influence of Falsified Position Data on Geographic Ad-hoc
  Eouting},''
\newblock in {\em Proceedings of the Second European conference on Security and
  Privacy in Ad-Hoc and Sensor Networks}, Berlin, Heidelberg, 2005, ESAS'05,
  pp. 102--112, Springer-Verlag.

\bibitem{jakobsson:VTC03}
M.~Jakobsson, S.~Wetzel, and B.~Yener,
\newblock ``{Stealth Attacks on Ad-hoc Wireless Networks},''
\newblock in {\em IEEE 58th Vehicular Technology Conference}, October 2003,
  vol.~3.

\bibitem{Sybil2011}
Tong Z., R.~Choudhury, Peng N., and K.~Chakrabarty,
\newblock ``{P2DAP-Sybil Attacks Detection in Vehicular Ad Hoc Networks},''
\newblock {\em IEEE Journal on Selected Areas in Communications}, vol. 29, no.
  3, pp. 582 --594, march 2011.

\bibitem{raya2006}
M.~Raya, P.~Papadimitratos, and J.-P. Hubaux,
\newblock ``Securing vehicular communications,''
\newblock {\em IEEE Wireless Communications}, vol. 13, no. 5, pp. 8 --15,
  october 2006.

\bibitem{bellman}
R.~Bellman,
\newblock ``{A Markovian Decision Process},''
\newblock {\em Journal of Mathematics and Mechanics}, vol. 6, no. 4, 1957.

\bibitem{getoor2007introduction}
L.~Getoor and B.~Taskar,
\newblock ``{Introduction to Statistical Relational Learning},'' The MIT Press,
  2007.

\bibitem{bertsekas2010approximate}
D.~Bertsekas,
\newblock ``{Approximate Policy Iteration: A Survey and Some New Methods},''
\newblock {\em Journal of Control Theory and Applications}, 2011.

\bibitem{sutton1998reinforcement}
R.~Sutton and A.~Barto,
\newblock ``{Reinforcement Learning},'' MIT Press, Volume 18, 1988.

\bibitem{busoniu2011cross}
L.~Busoniu, D.~Ernst, B.~De~Schutter, and R.~Babuska,
\newblock ``{Cross-Entropy Optimization of Control Policies with Adaptive Basis
  Functions},''
\newblock {\em IEEE Transactions on Systems, Man, and Cybernetics}, vol. 41,
  no. 1, pp. 196--209, 2011.

\bibitem{di2010adaptive}
D.~Di~Castro and S.~Mannor,
\newblock ``{Adaptive Bases for Reinforcement Learning},''
\newblock {\em Machine Learning and Knowledge Discovery in Databases}, pp.
  312--327, 2010.

\bibitem{yu2009basis}
H.~Yu and D.~Bertsekas,
\newblock ``Basis function adaptation methods for cost approximation in mdp,''
\newblock in {\em IEEE Symposium on Adaptive Dynamic Programming and
  Reinforcement Learning}, Nashville, TN, April 2009.

\bibitem{bertsekas-neuro}
D.~Bertsekas and J.~Tsitsiklis,
\newblock {\em {Neuro-Dynamic Programming}},
\newblock Athena Scientific, 1996.

\end{thebibliography}

\end{document}